\RequirePackage{amsmath}
\documentclass{llncs}  

\usepackage{cite}
\usepackage{amsfonts}
\usepackage{amssymb}
\usepackage{mathtools}
\usepackage{microtype}
\usepackage{stmaryrd}
\usepackage{xcolor}
\usepackage{epstopdf}
\usepackage{caption}
\usepackage[british,UKenglish]{babel}

\usepackage{hyperref}

\usepackage{tikz}
\usetikzlibrary{matrix,fit,calc,positioning,backgrounds,arrows,arrows.meta,patterns,decorations}
\usetikzlibrary{decorations.pathmorphing}
\tikzset{snake it/.style={decorate, decoration=snake}}

\tikzset{block/.style=%
  {draw,baseline,align=center,text width=2.5cm,minimum width=2.5cm,minimum height=.5cm}}
\tikzset{code/.style={block,fill=yellow!10,rounded corners}}
\tikzset{data/.style={block,fill=blue!10}}
\tikzset{arr/.style={draw,-{Stealth[round]},thick}}

\hypersetup{colorlinks,linkcolor=blue,citecolor=blue,urlcolor=blue}

\newcommand*\df\emph

\newcommand*\defeq{\mathrel{\coloneqq}}
\newcommand*\eval[1]{\left\llbracket#1\right\rrbracket}
\newcommand*\fA{\mathfrak{A}}

\newcommand*\liff{\leftrightarrow}
\newcommand*\limp{\to}
\newcommand*\lnand{\barwedge}
\newcommand*\sQbf{\fA_{\sf qbf}}

\newcommand*\Macyclic{\mathsf{acyclic}}
\newcommand*\Maej{\mathsf{AeJ}}
\newcommand*\Maj{\mathsf{AJ}}
\newcommand*\Mra{\mathsf{RA}}
\newcommand*\Mco{\mathsf{co}}
\newcommand*\Mcpp{\mathsf{CPP}}
\newcommand*\Mc{\mathsf{C}}
\newcommand*\Meq{\mathsf{eq}}
\newcommand*\Mf{\mathsf{F}}

\newcommand*\Mid{\mathsf{id}}
\newcommand*\Minj{\mathsf{inj}}
\newcommand*\Minv{\mathsf{inv}}
\newcommand*\Mirrefl{\mathsf{irrefl}}
\newcommand*\Mjr{\mathsf{JR}}
\newcommand*\Mj{\mathsf{J}}
\newcommand*\Mm{\mathsf{M}}
\newcommand*\Mrf{\mathsf{rf}}
\newcommand*\Mr{\mathsf{R}}
\newcommand*\Msc{\mathsf{SC}}
\newcommand*\Mseq{\mathsf{seq}}
\newcommand*\Ms{\mathsf{sub}}
\newcommand*\Mtc{\mathsf{TC}}
\newcommand*\Mtrans{\mathsf{trans}}
\newcommand*\Mv{\mathsf{V}}

\newcommand*\Yco{\mathit{Y}_{\mathit{co}}}
\newcommand*\Yhb{\mathit{Y}_{\mathit{hb}}}
\newcommand*\Yrf{\mathit{Y}_{\mathit{rf}}}

\renewcommand*\vec{\mathaccent"017E}

\newcommand*\ournote[3]{}
\newcommand*\RG[1]{\ournote{red}{RG}{#1}}

\newcommand*\co{\mathit{co}}
\newcommand*\rf{{\mathit{rf}}}

\newcommand*\hb{{\mathit{hb}}}
\newcommand*\po{{\mathit{po}}}
\newcommand*\ord{{\mathit{ord}}}
\newcommand*\cnf{{\mathit{conf}}}

\newcommand\evW[2]{{\rm W}_{#1}\,{#2}}
\newcommand\evR[2]{{\rm R}_{#1}\,{#2}}
\newcommand\evtlbl[1]{\mbox{#1:~}}

\title{PrideMM:\\ A Solver for Relaxed Memory Models}
\author{}
\institute{}
\titlerunning{Pride Memory Model Simulation}
\author{
  Simon Cooksey\inst{1}
  \and Sarah Harris\inst{1}
  \and Mark Batty\inst{1}
  \and Radu Grigore\inst{1}
  \and Mikol\'a\v{s}~Janota\inst{2}
}
\institute{
  University of Kent,
  \and
  IST/INESC-ID, University of Lisbon
}

\begin{document}
\maketitle
\vspace{-1cm}
\begin{abstract} 
  Relaxed memory models are notoriously delicate.
  To ease their study, several ad hoc simulators have been developed for axiomatic memory models.
  We show how axiomatic memory models can be simulated using a solver for~$\exists$SO\null.
  Further, we show how memory models based on event structures can be
  simulated using a solver for MSO\null.
  Finally, we present a solver for SO, built on top of QBF solvers.
\end{abstract} 
\section{Introduction}\label{sec:introduction} 
Understanding processor and language concurrency is an essential step
in building reliable systems.
Formal modelling and simulation have exposed
flaws~\cite{powerbug,hotspotunsound,gccbugs,amdbug} and led to
refinements~\cite{gpu,mathematizing} in the official descriptions of
concurrency in key languages and processors.
Current simulators rely on ad hoc
algorithms~\cite{herd,mathematizing,isamem} or SAT
solvers~\cite{memalloy}. However, flaws in existing language
concurrency models~\cite{thinairproblem} -- where one must account for
behaviour introduced through aggressive optimisation -- have led to a
new class of models~\cite{promising,Jeffrey:2016} that cannot be
simulated with previous ad hoc methods and fit awkwardly in the
limited language of SAT, making simulation unworkable.

This paper presents PrideMM, a tool that both simulates the more
intricate models of aggressively optimised concurrent languages and
replicates the functionality of previous tools. PrideMM identifies
second order (SO) logic as expressive enough to capture the wider set
of concurrency models, while restrictive enough to enable automatic
solving.
PrideMM uses a new checker, built above rapidly improving
\emph{quantified boolean formula} (QBF) solvers, that solves SO logic
formulas directly.

The following contributions underpin PrideMM:

\begin{enumerate}
\item we demonstrate simulation of existing models using a solver for~$\exists$SO\null,
\item we present a model checker for SO, built on top of QBF solvers, and
\item we simulate the Jeffrey and Riely model -- one of a new class of
  concurrency models for optimised concurrent languages -- using a
  solver for SO\null.
\end{enumerate}



\subsection{Modelling Relaxed Memory Models}
Processor speculation, memory-subsystem reordering and compiler
optimisations lead mainstream languages and processors to violate
\emph{sequential consistency}, a model of memory where accesses are
simply interleaved~\cite{lamportsc}. We say such systems exhibit \emph{relaxed
  concurrency}.  Relaxed concurrency is commonly described in an
\emph{axiomatic} model, where each program behaviour is represented as
graph of memory accesses, and a set of axioms filters forbidden execution graphs.

\begin{figure}[t]
\begin{center}
\begin{tikzpicture}[xscale=.9]
\node (a) at (4,0.8) {$\evtlbl{$a$}\evR{x}{1}$};
\node (b) at (4,0) {$\evtlbl{$b$}\evW{y}{1}$};
\node (c) at (6,0.8) {$\evtlbl{$c$}\evR{y}{1}$};
\node (d) at (6,0) {$\evtlbl{$d$}\evW{x}{1}$};
\draw[red,thick,->] (d) to [auto] node[above] {$\rf$} (a);
\draw[red,thick,->] (b) to [auto] node[below] {$\rf$} (c);
\draw[thick,->] (a) to [auto] node {$\po$} (b);
\draw[thick,->] (c) to [auto] node {$\po$} (d);
\node at (0,0.4) {
\begin{tabular}{c}
initially {\tt x = 0, y = 0}\\
\hline
\begin{tabular}{l @{\;} || @{\;} l}
{\tt r1 = x} & {\tt r2 = y} \\
{\tt if (r1 == 1)} & {\tt if (r2 == 1)} \\
{\tt \quad\quad \{y = 1\}} & {\tt \quad\quad \{x = 1\}}
\end{tabular} \\
\hline
{\tt r1 == 1, r2 == 1} allowed?
\end{tabular}
};
\node at (8.8,0.4) {$\text{acyclic}(\po \cup \rf)$
};
\end{tikzpicture}
\end{center}
\caption{\label{ex:axiomatic-LB}
LB+ctrl, an axiomatic execution of it, and an axiom that forbids it.}
\end{figure}
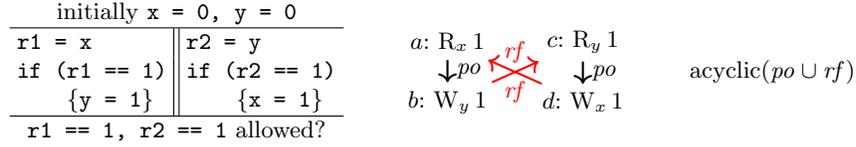

Herd is a simulator of axiomatic models that has been used extensively
to model processor, GPU, and language concurrency~\cite{herd}. In Herd,
the model is expressed as a predicate on execution graphs, written in
the propositional relation calculus, and recorded in a {\tt .cat}
file. Figure~\ref{ex:axiomatic-LB} presents \emph{load buffering with
  control dependencies} (LB+ctrl), a small program called a
\emph{litmus test} constructed to probe for a single relaxed
behaviour, together with an execution graph and an axiom as it would
appear in a {\tt .cat} file. LB+ctrl consists of two parallel threads
that read {\tt x} (or {\tt y}) and then conditionally write {\tt y}
(or {\tt x}), with {\tt x} and {\tt y} initialised to 0. The outcome
1/1 represents a relaxed behaviour, and is allowed in particular by
the current C++ standard, but forbidden under the SC, x86, Power and
ARM models. The graph of Figure~\ref{ex:axiomatic-LB} presents the
execution in question, with memory reads and writes as vertices
(eliding the initialisation) and edges representing program order
($\po$) and the writes that each read reads from ($\rf$). The axiom of
Figure~\ref{ex:axiomatic-LB} forbids the outcome 1/1 as the
corresponding execution contains a cycle in $\po \cup \rf$. The SC,
x86, Power and ARM models each include a variant of this axiom,
all forbidding 1/1.

Herd uses an ad hoc algorithm for judging whether an execution is allowed. Its
performance is surpassed by the Memalloy~\cite{memalloy} tool built above
SAT-based Alloy, so it is clear that the judgement of axiomatic models can be
expressed as a SAT problem. Unfortunately, not all memory models fit the
axiomatic paradigm.

\begin{figure}[t]
\begin{center}
\begin{tikzpicture}[xscale=.9]
\node at (0,1) {
\begin{tabular}{c}
initially {\tt x = 0, y = 0}\\
\hline
\begin{tabular}{l @{\;} || @{\;} l}
{\tt r1 = x} & {\tt r2 = y} \\
{\tt if (r1 == 1)} & {\tt if (r2 == 1)} \\
{\tt \quad\quad \{y = 1\}} & {\tt \quad\quad \{x = 1\}} \\
& {\tt else} \\
& {\tt \quad\quad \{x = 1\}}
\end{tabular} \\
\hline
{\tt r1 == 1, r2 == 1} allowed?
\end{tabular}
};
\node(i) at (6.5,2) {$\mathit{Init}$};
\node (a) at (3.5,0.8) {$\evtlbl{$a$}\evR{x}{0}$};
\node (b) at (5.5,0.8) {$\evtlbl{$b$}\evR{x}{1}$};
\node (c) at (5.5,0) {$\evtlbl{$c$}\evW{y}{1}$};
\node (d) at (7.5,0.8) {$\evtlbl{$d$}\evR{y}{0}$};
\node (e) at (7.5,0) {$\evtlbl{$e$}\evW{x}{1}$};
\node (f) at (9.5,0.8) {$\evtlbl{$f$}\evR{y}{1}$};
\node (g) at (9.5,0) {$\evtlbl{$g$}\evW{x}{1}$};
\draw[thick,->,bend right] (i) to [auto] node[above left] {$\ord$} (a);
\draw[thick,->] (i) to [auto] node[above left] {$\ord$} (b);
\draw[thick,->] (i) to [auto] node[above right] {$\ord$} (d);
\draw[thick,->,bend left] (i) to [auto] node[above right] {$\ord$} (f);
\draw[thick,->] (b) to [auto] node {$\ord$} (c);
\draw[thick,->] (d) to [auto] node {$\ord$} (e);
\draw[thick,->] (f) to [auto] node[left] {$\ord$} (g);
\draw[red,thick,snake it] (a) to [auto] node[above] {$\cnf$} (b);
\draw[red,thick,snake it] (d) to [auto] node[above] {$\cnf$} (f);
\end{tikzpicture}
\end{center}
\caption{\label{ex:eventstructure-LBfalse}
LB+false-ctrl and the corresponding event structure.}
\end{figure}
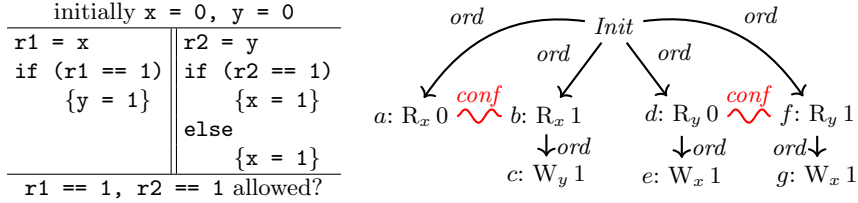

\paragraph{Axiomatic models do not fit optimised languages.} Languages
like C++ and Java perform dependency-removing optimisations that
complicate their memory models. For example, the second thread of the
LB+false-dep test in Figure~\ref{ex:eventstructure-LBfalse} can be
optimised using common subexpression elimination to {\tt r2=y;
  x=1;}. On ARM and Power, this optimised code may be reordered,
permitting the relaxed outcome 1/1, whereas the syntactic dependency
of the original would make 1/1 forbidden. It is common practice to use
syntactic dependencies to enforce ordering on hardware, but at the
language level the optimiser removes these \emph{fake} dependencies.

The C++ standard is flawed because it describes an axiomatic language
model that cannot draw a distinction between the executions leading to
outcome 1/1 in LB+dep and LB+false-dep: the details of other branches
of control flow have been stripped and they have precisely the same
vertices and edges~\cite{thinairproblem}

\paragraph{Event structures capture the necessary information.}
A new class of models aims to fix this by ordering only real
dependencies~\cite{promising,Jeffrey:2016,bubbly}. With a notable
exception~\cite{promising}, these models are based on \emph{event
  structures}, where all paths of control flow are represented in a
single graph. Figure~\ref{ex:eventstructure-LBfalse} presents the
event structure for LB+false-deps. Program order is captured by the
$\ord$ relation. \emph{Conflict}, the $\cnf$ edge, links events where
only one can occur in an execution (the same holds for their
$\ord$-successors). For example, on the left-hand thread, the load of
{\tt x} can result in a read of value 0 (event $a$) or a read of value
1 (event $b$), but not both.  Conversely, two subgraphs unrelated by
$\ord$ or $\cnf$, e.g. $\{a,b,c\}$ and $\{d,e,f,g\}$, represent two
threads in parallel execution.

It should be clear from the event structure in
Figure~\ref{ex:eventstructure-LBfalse} that regardless of the value
read from {\tt y} in the right-hand thread, there is a write to {\tt
  x} of value 1, i.e. the apparent dependency from the load of {\tt y}
is fake and could be optimised away. Memory models built above event
structures can recognise this pattern and permit relaxed execution.

\paragraph{The Jeffrey and Riely model.}
Jeffrey and Riely proposed a concurrency model (henceforth referred to
as J+R) built above event structures that correctly identifies fake
dependencies~\cite{Jeffrey:2016}.  Conceptually, the model is related
to the Java memory model~\cite{javamm}: in both, one constructs an
execution stepwise, adding only memory events that can be
\emph{justified} from the previous steps. The sequence captures a
causal order and prevents cycles that could lead to thin-air
values. While Java is too strong, the J+R model allows writes that
have fake dependencies on a read to be justified before that read. To
do this, the model recognises confluence in the program structure:
regardless of the execution path, the write will always be made. This
search across execution paths involves alternation of quantification
that current ad hoc and SAT-based tools cannot efficiently simulate.
The problem is amenable to the new breed of QBF solvers.

\subsection{Solvers}
The late 90's brought about a surge of practical applications of \emph{SAT solvers}~\cite{silva99,SatHandbook}.
QBF provides a more expressive language and therefore less burden on the modeler but it is also inherently harder.
Indeed, QBF is PSPACE-complete, whereas SAT is ``only'' NP-complete.
Initially QBF solving mainly focused on adapting SAT techniques to quantifiers~\cite{zhang-iccad02}.
In the last decade, however, there has been a prolific activity in the field leading to several independent paradigms.
There has been a  remarkable progress in the area almost each year~\cite{klieber-sat10,GoultiaevaB10,janota-sat11,janota-sat12,goultiaeva-date13,Gelder-cp13,BalabanovSAT16,malik2qbf,janota-ijcai15,rabe-fmcad15,LonsingSAT16,tentrup-sat16,rabe-sat16,peitl-sat17}.
This  evolution has also been  traced by the yearly QBF competitions~\cite{qbfeval}, see also~\cite{marin-fi16}.
These improvements suggest that it may be beneficial to integrate modern QBF technology into formal verification tools.

We highlight the algorithm \emph{RAReQS}~\cite{janota-sat12,janota-ai16} with its recent improvements~\cite{janota-aaai18}.
The algorithm has exhibited highly competitive performance in formulas coming from practical applications and with small number of quantifier
levels.  Hence, RAReQS is a natural candidate for the problems targeted in this paper.
Nevertheless, other solvers are also included in the evaluation (see~\autoref{sec:evaluation}).

From practical perspective, it is important to mention the input format to QBF solvers.
Unlike in SAT, CNF input has been observed as extremely harmful to QBF solving~\cite{zhang-aaai06,selman-aaai05,janota-epia17}.
This has been reflected by recent efforts to promote solvers that accept a circuit-like format \emph{QCIR}~\cite{klieber-bnp16}.
Hence, QBF solvers can be classified  according to which of the two inputs they support.
During the experimental evaluation we have observed that the circuit-based solvers dramatically outperform the CNF-based ones
(see \autoref{sec:evaluation}).

We  should mention that there are other tools  dedicated to automated solving
in higher-order logic.  Namely \emph{higher order model
finders}~\cite{blanchette-itp10} or \emph{automated higher order theorem
provers}~\cite{brown-ijcar12}. Even though one could encode the problems
considered in this paper into those tools, their ultimate focus
are mainly mathematical theorems. Hence, applying these tools to our
problems would likely lead to poor performance: a scenario of a using a sledgehammer to crack
a nut.


\section{Overview}\label{sec:overview} 

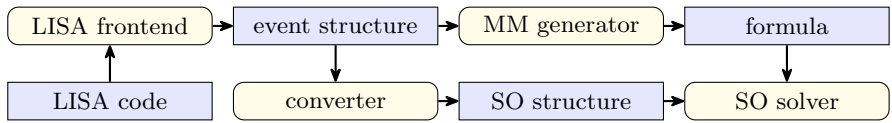
\begin{figure}\centering
\begin{tikzpicture}[on grid,node distance=1cm and 3cm]
\node[data] (lisa-c) {LISA code};
\node[code,above=of lisa-c] (lisa-f) {LISA frontend};
\node[data,right=of lisa-f] (es) {event structure};
\node[code,right=of es] (mmgen) {MM generator};
\node[data,right=of mmgen] (phi) {formula};
\node[code,below=of es] (es-to-so) {converter};
\node[data,right=of es-to-so] (str) {SO structure};
\node[code,right=of str] (so-solver) {SO solver};

\draw[arr] (lisa-c) -- (lisa-f);
\draw[arr] (lisa-f) -- (es);
\draw[arr] (es) -- (mmgen);
\draw[arr] (mmgen) -- (phi);
\draw[arr] (es) -- (es-to-so);
\draw[arr] (es-to-so) -- (str);
\draw[arr] (str) -- (so-solver);
\draw[arr] (phi) -- (so-solver);
\end{tikzpicture}
\caption{
  From a LISA test case to a Y/N answer, given by the SO solver.
}\label{fig:architecture}
\end{figure}

Figure \ref{fig:architecture} shows the architecture of our memory-model simulator.
The input is LISA code, and the output is a yes\slash no answer.
LISA is a programming language that has been designed for studying memory models~\cite{lisa}.
This language enables writing  multi-threaded programs and asking questions about whether certain behaviours are allowed.
Using LISA as our input format enables a comparison with the state-of-the-art memory-model simulator Herd~\cite{herd}.
The LISA frontend produces an event structure~\cite{Winskel:1987}.
Any event structure is trivially representable as a SO logic structure, so the conversion is simple.
The MM~generator (memory-model generator) produces a SO formula.
We have a few interchangeable MM~generators (\autoref{sec:mm}).
For some memory models (\autoref{sec:mm:sc}, \autoref{secgc}, \autoref{sec:mm:cpp}), which Herd can handle as well, the formula is in fact fixed and does not depend at all on the event structure.
For other memory models (such as~\autoref{sec:mm:jr}), the
MM~generator might need to look at certain characteristics of the
event structure (such as its size).
Finally, both the second-order structure and the second-order formula are fed into a solver, which effectively simulates the program under the memory-model, and gives a verdict.

We build on prior work from two different areas -- relaxed memory models, and SAT\slash QBF solving:
the LISA frontend comes from the Herd memory-model simulator~\cite{herd}, the MM~generators implement memory models that have been previously proposed~\cite{repairingsc,Jeffrey:2016}, and the SO solver is based on a state-of-the-art QBF solver~\cite{janota-aaai18}.
Our main contribution to the area of relaxed memory models is that we widen the class of memory models that can be efficiently simulated.
Our main contribution to SAT/QBF solving is that we widen the applicability of such tools.

Applying SAT technology to simulate memory models has been tried before~\cite{memalloy}.
But, although it did lead to performance improvements, it did not widen the class of models that can be efficiently simulated.
We are able to do so because of a key insight: relational second-order logic represents a sweet-spot in the design space.
On the one hand, it is expressive enough such that encoding memory models is natural.
On the other hand, it is simple enough such that it can be solved efficiently, using emerging QBF technology.

Consider for example the sequentially consistent memory model.
It is often described by saying that there exists a reads-from relation~$\rf$ and a coherence order~$\co$ such that the transitive closure of $\rf\cup\co\cup(\rf^{-1};\co)\cup\po$ is acyclic.
Here, $\po$ is the (fixed) program-order relation, and it is understood that $\co$~and~$\rf$ satisfy certain further axioms.
In our setting, we describe the sequentially consistent model as follows.
We represent $\rf$~and~$\co$ by existentially-quantified SO arity-$2$ variables $\Yco$~and~$\Yrf$, respectively.
For example, to say $(x,y)\in\co$, we use the formula~$\Yrf(x,y)$.
The program order $\po$ is represented by an interpreted arity-$2$ symbol~$<$.
Then, the SO formula that represents $\rf\cup\co\cup(\rf^{-1};\co)\cup\po$ is
\begin{align}
  \Mr(y,z) \;&\defeq\;
    \Yrf(y,z)
    \lor
    \Yco(y,z)
    \lor
    \exists x\, \bigl( \Yrf(x,z) \land \Yco(x,y) \bigr)
    \lor
    (y < z)
\end{align}
The definition from above should be interpreted as a macro expansion rule: the left-hand side $\Mr(y,z)$ is a macro that expands to the formula on right-hand side.
To require that the transitive closure of $\Mr$ is acyclic we require that there exists a relation that includes $\Mr$, is transitive, and irreflexive:
\begin{align}
  \exists Z\, \bigl(
    \Ms^2(\Mr, Z)
    \land
    \Mtrans(Z)
    \land
    \Mirrefl(Z)
  \bigr)
\end{align}
The macros $\Ms^2$, $\Mtrans$, $\Mirrefl$ are defined as one would expect.
For example, $\Ms^2(P,Q)$, which says that the arity-$2$ relation $P$ is included in the arity-$2$ relation~$Q$, is $\forall xy\,\bigl(P(x,y)\land Q(x,y)\bigr)$.
In short, the translation from the usual formulation of memory-models into the SO logic encoding that we propose is natural and almost automatic.
In~\autoref{sec:mm}, we describe this translation in detail for $4$~memory models.
One of these models~(\autoref{sec:mm:jr}) illustrates that the translation is not entirely automatic:
some care is required to skirt exponential blowup.

To represent programs and their behaviours uniformly for all memory models, we use event structures.
These have the ability to represent an overlay of potential executions.
Some memory-models require reasoning about several executions at the
same time: this is a salient feature of the J+R memory model. 

Once we have the program and its behaviour represented as a logic structure~$\fA$ and the memory model represented as a logic formula~$\phi$, we ask whether the structure satisfies the formula, written~$\fA \models \phi$.
In other words, we have to solve a model-checking problem for second-order logic, which reduces to QBF solving because the structure $\fA$ is finite.
As a foretaste, consider the SO formula
\begin{align}
  \exists X\, \left(
  \begin{aligned}
    &\forall xy\, \bigl((\mathtt{ord}(x,y) \land X(y)) \limp X(x) \bigr)
    \land
    \\&\quad
    \forall xy\, \bigl((X(x)\land X(y)) \limp \lnot \mathtt{conflict}(x,y)\bigr)
  \end{aligned}
  \right)
\label{eq:first-so-example}
\end{align}
which asks if there exists an execution~$X$ that is downward closed with respect to the order~$\ord$ and does not contain conflicting events.
We wish to evaluate this formula on a structure $\fA$ defined by
\begin{align}
A=\{1,2,3\}
\qquad
\mathtt{ord}^\fA \defeq \{(1,2),\, (1,3)\}
\qquad
\mathtt{conflict}^\fA \defeq \{ (2,3) \}
\end{align}
This structure contains three events that are partially ordered (with $1$~coming first).
Events $2$~and~$3$ are conflicting.
The QBF question we ask is the following:
\begin{align}
\forall\mathsf{x}_1\mathsf{x}_2\mathsf{x}_3\, \bigl(
  (\mathsf{x}_2 \limp \mathsf{x}_1)
  \land
  (\mathsf{x}_3 \limp \mathsf{x}_1)
  \land
  \lnot(\mathsf{x}_2 \land \mathsf{x}_3)
\bigr)
\label{eq:first-qbf-example}
\end{align}
To represent the arity-$1$ second-order variable~$X$ we introduced $3$~(Boolean) QBF variables $\mathsf{x}_1, \mathsf{x}_3, \mathsf{x}_3$.
In general, an arity-$k$ second-order variable is encoded into $|A|^k$ QBF variables.
The first order quantifiers ($\forall xy$) disappeared altogether, because they were expanded.
The relation names $\mathtt{ord}$ and $\mathtt{conflict}$ do not appear anymore either, because, once we fixed $x$~and~$y$, we could replace them by true\slash false constants that were simplified away.
For example, $\mathtt{ord}(2,3)$ was replaced by `false', which was simplified away; and $\mathtt{ord}(1,2)$ was replaced by `true', which was also simplified away.

Observe that \eqref{eq:first-qbf-example} is in fact a SAT instance.
This is because the formula~\eqref{eq:first-so-example} does not contain universal second-order variables.
When such universal variables are present, they give rise naturally to universal QBF variables.

It is well known that SO finite model-checking can be reduced to QSAT\null.
However, in practice it is important to know how the reduction is done.
We give the details of our reduction in~\autoref{sec:qbf}.
We have implemented this reduction twice, independently.
One implementation is built-in the SO solver and optimised; the other implementation is an optional backend for the MM generators.
Having two implementations that agree increases our confidence that they are correct.
Further, the MM generator backend produces formulas in the QCIR format, which can be solved using multiple QBF solvers.

\smallskip

We illustrate the generality of our approach by implementing the MM~generator component for $4$~memory models, including one that cannot be handled by existing simulators.
These $4$~MM~generators are implemented on top of an OCaml API that provides combinators such as $\Ms^2$, $\Mtrans$, and~$\Mirrefl$.
Since this API has $4$~users, we believe it is reusable.

Three of the four memory models we described could be described in the CAT language~\cite{cat}, but not J+R\null.
As future work, we aim to extend the CAT language, and implement a generic MM generator that can handle this extended CAT\null.

\section{Preliminaries}\label{sec:preliminaries} 

The standard problem solved by simulators is to decide whether a given program behaviour is allowed by a given memory model.
The standard model checking problem is to decide whether a given structure~${\mathfrak A}$ satisfies a formula~$\phi$, written ${\mathfrak A} \models \phi$.
We will describe program behaviours by relational structures~$\mathfrak A$, and memory models by second-order formulas~$\phi$.

We now recall standard definitions~\cite{Libkin:2004}.
A (finite, relational) \df{vocabulary}~$\sigma$ is a finite collection of \df{constant symbols} ($a$, $b$,~\dots) together with a finite collection of \df{relation symbols} ($Q$, $R$,~\dots).
A (finite, relational) \df{structure~$\fA$ over vocabulary~$\sigma$} is a tuple $\langle A, b^\fA, c^\fA, \ldots, Q^\fA, R^\fA, \ldots \rangle$ where
  $A$~is a finite set called \df{universe} with several distinguished elements $a^\fA,b^\fA,\ldots$ and relations $Q^\fA,R^\fA,\ldots$
To simplify the presentation, we will assume that the universe $A$ is $\{a_1^\fA,\ldots,a_n^\fA\}$, and that the constant symbols include $a_1,\ldots,a_n$, which denote the elements of the universe.
For each distinguished relation such as~$Q^\fA$, there is a~$k$ such that $Q^\fA \subseteq A^k$; we say that $k$ is the \df{arity} of~$Q^\fA$.
We assume a countable set of \df{first-order variables} ($x$, $y$,~\dots);
for each arity $k>0$, we assume a countable set of \df{second-order variables} ($X^k$, $Y^k$,~\dots).
In particular, we think of the arity as being part of the variable name, and we single it out only when necessary.
A \df{variable}~$\alpha$ is a first-order variable or a second-order variable;
a \df{term}~$t$ is a first-order variable or a constant symbol;
a \df{predicate}~$P^k$ is a second-order variable or a relation symbol.
A (second-order) \df{formula}~$\phi$ is defined inductively:
(a)~if $P^k$ is a predicate and $t_1,\ldots,t_k$ are terms, then $P^k(t_1,\ldots,t_k)$ is a formula;
(b)~if $\phi_1$~and~$\phi_2$ are formulas, then $\phi_1 \lnand \phi_2$ is a formula;
(c)~if $\alpha$~is a variable and $\phi$~is a formula, then $\forall\alpha\,\phi$ and $\exists\alpha\,\phi$ are formulas.
Other boolean connectives can be desugared into logical-not-and~$\lnand$.

Assume a structure $\fA$ over universe~$A$, a formula~$\phi$, an environment~$\gamma$ that binds the free first-order variables of~$\phi$ to elements of~$A$, and an environment~$\Gamma$ that binds the free SO variables of~$\phi$ to subsets of~$A^k$, where $k$~is the arity.
We use the notation $\gamma[x\mapsto a^\fA]$ and $\Gamma[x\mapsto R^\fA]$ to extend environments, which we define as $\gamma[x \mapsto a^\fA](y) \:\defeq\:a^\fA$ when $y = x$ and $\gamma(y)$ otherwise. Similar for $\Gamma$.

  
We let the first-order empty environment~$\epsilon$ map constant symbols to their respective constants $\epsilon(a)\defeq a^\fA$, and we let the second-order empty environment~$E$ map relation symbols to their respective relations $E(R) \defeq R^\fA$.
With these conventions, we interpret formulas over structures by defining the judgement $\fA \models \phi [\gamma,\Gamma]$ as follows:
\begin{align*}
\fA &\models P(t_1,\ldots,t_k)[\gamma,\Gamma]
  &&\text{iff}
  &&\bigl(\gamma(t_1),\ldots,\gamma(t_k)\bigr) \in \Gamma(P)
\\
\fA &\models (\phi_1 \lnand \phi_2)[\gamma,\Gamma]
  &&\text{iff}
  &&\text{not both $\fA \models \phi_1[\gamma,\Gamma]$ and $\fA\models\phi_2[\gamma,\Gamma]$}
\\
\fA &\models (\forall x\, \phi)[\gamma,\Gamma]
  &&\text{iff}
  &&\text{$\fA \models \phi\bigl[\gamma[x\mapsto a^\fA],\Gamma\bigr]$
    for all $a^\fA\in A$}
\\
\fA &\models (\exists x\, \phi)[\gamma,\Gamma]
  &&\text{iff}
  &&\text{$\fA \models \phi\bigl[\gamma[x\mapsto a^\fA],\Gamma\bigr]$
    for some $a^\fA\in A$}
\\
\fA &\models (\forall X^k\, \phi)[\gamma,\Gamma]
  &&\text{iff}
  &&\text{$\fA \models \phi\bigl[\gamma,\Gamma[X^k\mapsto R^\fA]\bigr]$
    for all $R^\fA\subseteq A^k$}
\\
\fA &\models (\exists X^k\, \phi)[\gamma,\Gamma]
  &&\text{iff}
  &&\text{$\fA \models \phi\bigl[\gamma,\Gamma[X^k\mapsto R^\fA]\bigr]$
    for some $R^\fA\subseteq A^k$}
\end{align*}

\RG{Define FV\null.}
The notation $\fA\models\phi$ is a shorthand for $\fA\models\phi[\epsilon,E]$.
A formula with no free variables is called a \df{sentence}.
\RG{Use `sentence' where appropriate.}
For a formula~$\phi$ whose free variables are $\vec{\alpha}$, both $\exists\vec{\alpha}\,\phi$ and $\forall\vec{\alpha}\,\phi$ are sentences.
We say that $\phi$~is \df{satisfiable} when there exists a structure~$\fA$ such that $\fA \models \exists\vec{\alpha}\, \phi$;
we say that $\phi$~is \df{valid} when for all structures~$\fA$ we have $\fA \models \forall\vec{\alpha}\, \phi$.

The logic defined so far is known as~SO\null.
If we require that all quantifiers over second-order variables are existentials, we obtain a fragment known as $\exists$SO (existential second-order).
If we require that all second-order variables have arity~$1$, we obtain a fragment known as MSO (monadic second-order).
If we make both requirements, the fragment is called $\exists$MSO\null.

\begin{figure}[t]
  \centering
\[
\begin{array}{@{}r@{\;\defeq\;}l@{\qquad}r@{\;\defeq\;}l@{}}
\Ms^k(P^k,Q^k) &
  \forall \vec{x}\, \bigl( P^k(\vec{x}) \limp Q^k(\vec{x}) \bigr)
&
\Mid(x,y) & (x=y)
\\
\Meq^k(P^k,Q^k) &
  \forall \vec{x}\, \bigl( P^k(\vec{x}) \liff Q^k(\vec{x}) \bigr)
&
\Minj(P) &
  \Ms^2\bigl(\Mseq(P,\Minv(P)), \Mid\bigr)
\\
\Mirrefl(P) & \forall x\, \lnot P(x,x)
&
\Mseq(P,Q)(x,z) & \exists y\, \bigl( P(x,y) \land Q(y,z) \bigr)
\\
\Minv(P)(x,y) & P(y,x)
&
\Mtrans(P) &
  \Ms^2\bigl(\Mseq(P,P),P\bigr)
\end{array}
\]
\[
\begin{array}{@{}r@{\;\defeq\;}l@{}}
\Macyclic(P) &
  \exists X\, \bigl(
    \Ms^2(P,X)
    \land
    \Mtrans(X)
    \land
    \Mirrefl(X)
  \bigr)
\\
\Mtc_0(\Mr) & \Meq^1
\\
\Mtc_{n+1}(\Mr)(P^1,Q^1) &
  \Meq^1(P^1,Q^1)
  \lor
  \exists X^1\, \bigl(
    \Mr(P^1,X^1)
    \land
    \Mtc_n(\Mr)(X^1,Q^1)
  \bigr)
\end{array}
\]
\caption{
  Combinators used to build SO formulas.
  By convention, all quantifiers that occur on the right-hand side of the definitions above are over fresh variables.
  Above, $P$~and~$Q$ are arity-$2$ predicates, $P^k$~and~$Q^k$ are arity-$k$ predicates, $x$~and~$y$ are first-order variables, and $\Mr$ is a combinator.
}\label{fig:combinators}
\end{figure}

\paragraph{Combinators.}
In what follows, we shall be describing some rather large SO formulas.
To do so concisely, we shall utilise the combinators from Figure~\ref{fig:combinators}.
All combinators are typeset in \textsf{sf-fonts}.

Let us discuss two of the more interesting combinators: $\Macyclic$ and~$\Mtc$.
A relation $P$ is acyclic if it is included in a relation that is transitive and irreflexive.
We remark that the definition of $\Macyclic$ is carefully chosen: even slight variations can have a strong influence on the runtime of solvers~\cite{acyclicity}.
The combinator $\Mtc$ for bounded transitive closure is interesting for another reason: it is higher-order.
By way of example, let us illustrate its application to the subset combinator~$\Ms^1$.
\begin{small}
\begin{align*}
&\Mtc_1(\Ms^1)(P,Q)
\\&\qquad=
  \Meq^1(P,Q)
  \lor
  \exists X\, \bigl(
    \Ms^1(P,X)
    \land
    \Mtc_0(\Ms^1)(X,Q)
  \bigr)
\\&\qquad=
  \left\{
  \begin{aligned}
  &\forall x_1\, \bigl(P(x_1)\liff Q(x_1)\bigr)
  \lor
  \\
  &\quad\exists X\, \bigl(
    \forall x_2\, \bigl( P(x_2) \limp X(x_2) \bigr)
    \land
    \Meq^1(X,Q)
  \bigr)
  \end{aligned}
  \right.
\\&\qquad=
  \left\{
  \begin{aligned}
  &\forall x_1\, \bigl(P(x_1)\liff Q(x_1)\bigr)
  \lor
  \\
  &\quad\exists X\, \bigl(
    \forall x_2\, \bigl( P(x_2) \limp X(x_2) \bigr)
    \land
    \forall x_3\, \bigl( X(x_3) \liff Q(x_3) \bigr)
  \bigr)
  \end{aligned}
  \right.
\end{align*}
\end{small}%
In the calculation above, $P$, $Q$ and~$X$ have arity~$1$.
In what follows, we freely use the combinators from Figure~\ref{fig:combinators} and, occasionally, we define some that are specific to a memory model.


\section{Memory Models}\label{sec:mm} 
%
%
In this section, we show that many memory models can be expressed conveniently in second-order logic.
Before diving into the details of memory models, let us first discuss briefly the representation we use for programs and their behaviours; namely, event structures.
We first describe the theory of event structures (vocabulary and axioms), followed by some useful definitions and notational conventions.
We do not describe how event structures are obtained from programs; for that, we refer the reader to~\cite{Jeffrey:2016}.

\paragraph{Vocabulary.}

A memory model decides if a program is allowed to have a certain behaviour.
We shall formulate this question as a model checking question, $\fA \models \phi$.
The vocabulary of~$\fA$ consists of the following symbols:
\begin{itemize}
\item arity~1: $\mathtt{final}$, $\mathtt{read}$, $\mathtt{write}$
\item arity~2: $\mathtt{conflict}$, $\mathtt{justifies}$, $\mathtt{sloc}$, $\le$, $=$
\end{itemize}
The symbol~$=$ always denotes the identity relation on events, $\{\,(x,x)\mid x\in A\,\}$.
%
%
The symbol~$\le$ corresponds to program order; we have $x \le y$ when
events $x$~and~$y$ come from program statements that are ordered in
the program text.
We have $\mathtt{justifies}(x,y)$ when $x$~reads the value that
$y$~wrote, to the same memory location.
%
We have $\mathtt{conflict}(x,y)$ when events $x$~and~$y$ cannot belong to the same execution; for example, events $x$~and~$y$ may model the same read-statement but for different values that are being read.
The sets $\mathtt{read}$ and $\mathtt{write}$ classify events in the obvious way.
We have $\mathtt{sloc}(x,y)$ when $x$~and~$y$ are access the same memory location.

The symbol $\mathtt{final}$ is not a standard component of event structures.
We will make use of it to identify the set of executions that exhibit a behaviour of interest.

\paragraph{Axioms.}

The theory of event structures is defined by the following axioms:
\begin{align}
\fA &\models \forall x\,
  \bigl(\lnot\mathtt{read}(x) \lor \lnot\mathtt{write}(x)\bigr)
\\
\fA &\models \forall xy\, \bigl(
  \mathtt{justifies}(x,y)
  \limp
  \bigl(\mathtt{write}(x) \land \mathtt{read}(y)\bigr)
\bigr)
\\
\fA &\models \forall xy\, \bigl(
  \mathtt{conflict}(x,y) \liff \mathtt{conflict}(y,x)
\bigr)
\\
\fA &\models \forall x\, \lnot \mathtt{conflict}(x,x)
\\
\fA &\models \forall xyz\, \bigl(
  \bigl(\mathtt{conflict}(x,y) \land (y\le z)\bigr)
  \limp
  \mathtt{conflict}(x,z)
\bigr)
\\
\fA &\models \forall xyz\, \bigl(
  \bigl(\mathtt{conflict}(x,y) \land (z < y)\bigr)
  \limp
  (z < x)
\bigr)
\\
\fA &\models \forall xyz\, \left(
  \begin{aligned}
  &\bigl(
    \mathtt{conflict}(x,y) \land \mathtt{conflict}(y,z)
  \bigr)
  \\&\quad\limp
  \bigl(
    \mathtt{conflict}(x,z) \lor (x=z)
  \bigr)
  \end{aligned}
\right)
\end{align}
Intuitively, conflicts can first occur when an event~$x$ is immediately followed in program-order by two events $y_1$~and~$y_2$ which are incomparable to each-other; and once a conflict occurs it propagates to subsequent events.
Furthermore, conflict is irreflexive, and becomes transitive when unioned with the identity relation.

Currently, our SO solver has no knowledge of the theory of event structures, so it does not exploit the axioms from above.
But, it can check that the structures~$\fA$ we produce satisfy the axioms, as they should.

\paragraph{Configurations and Executions.}

We distinguish two types of sets of events.
A \df{configuration} is a set of events that contains no conflict and is downward closed with respect to~$\le$; that is, $X$~is a configuration when $\Mv(X)$ holds, where the $\Mv$~combinator is defined by
\begin{align}
  \Mv(X) \;\defeq\;
\left\{
\begin{aligned}
    &\forall x \forall y\,
      \Bigl( \bigl(X(x) \land X(y)\bigr) \limp \lnot \mathtt{conflict}(x,y) \Bigl)
    \\&\quad{}\land \forall y\, \Bigl( X(y) \limp \forall x\, \bigl( (x\le y)\limp X(x)\bigr) \Bigr)
\end{aligned}
\right.
\end{align}

We say that a configuration $X$ is an \df{execution of interest} when every final event is either in~$X$ or in conflict with an event in~$X$; that is, $X$~is an execution of interest when $\Mf(X)$ holds, where the $\Mf$~combinator is defined by
\begin{align}
  \Mf(X) \;\defeq\;
  \Mv(X) \land
  \forall x\, \left(
\begin{aligned}
  &\bigl(\mathtt{final}(x) \land \lnot X(x)\bigr) \limp
  \\&\quad
  \exists y\, \bigl(
    \mathtt{conflict}(x,y)
    \land \mathtt{final}(y)
    \land X(y) \bigr)
\end{aligned}
\right)
\end{align}
Intuitively, we shall put in $\mathtt{final}$ all the maximal events (according to~$\le$) for which registers have the desired values.

\paragraph{Notations.}

In the formulas below, $X$ will stand for a configuration, which may be the execution of interest.
Variables $\Yrf$, $\Yco$, $\Yhb$ and so on are used to represent the relations that are typically denoted by $\rf$, $\co$, $\hb$, \dots
Thus, $X$~has arity~$1$, while $\Yrf,\Yco,\dots$ have arity~$2$.

\smallskip
In what follows, we present four memory models: sequential consistency (\autoref{sec:mm:sc}), release--acquire (\autoref{secgc}), C++~(\autoref{sec:mm:cpp}), and J+R~(\autoref{sec:mm:jr}).
The first three can be expressed in $\exists$SO (and in first-order logic).
The last one uses both universal and existential quantification over sets.
For each memory model, we shall see their encoding in second-order logic.

\subsection{Sequential Consistency}\label{sec:mm:sc} 

The sequential consistency memory model is the oldest and the least relaxed we consider.
Intuitively, this model allows all interleavings of threads, and nothing else.
%
It is described by the following SO~sentence:
\begin{align*}
\Msc \defeq
  \exists X\Yco\Yrf\, \bigl(
    \Mf(X) \land \Mco(X,\Yco) \land \Mrf(X,\Yrf) \land \Macyclic(\Mr(\Yco,\Yrf))
  \bigr)
\label{eq:mm:sc}
\end{align*}
Intuitively, we say that there exists a coherence order relation~$\Yco$ and a reads-from relation~$\Yrf$ which, when combined in a certain way, result in an acyclic relation $\Mr(\Yco,\Yrf)$.
The formula $\Mco(X,\Yco)$ says that $\Yco$~satisfies the usual axioms of a coherence order with respect to the execution~$X$;
and the formula $\Mrf(X,\Yrf)$ says that $\Yrf$~satisfies the usual axioms of a reads-from relation with respect to the execution~$X$.
Moreover, the formula $\Mf(X)$ asks that $X$~is an execution of interest, which results in registers having certain values.
\begin{small}
\begin{align}
  \Mco(X,\Yco) &\defeq
    \forall xy\, \left(
    \begin{aligned}
      &\bigl(
        X(x) \land X(y) \land
        \mathtt{write}(x) \land \mathtt{write}(y) \land \mathtt{sloc}(x,y) \land (x\ne y)
      \bigr)
      \\&\quad
      \liff
      \bigl(\Yco(x,y) \lor \Yco(y,x)\bigr)
    \end{aligned}
    \right)
\\
  \Mrf(X,\Yrf) &\defeq
  \left\{
  \begin{aligned}
    &\Minj(\Yrf)
    \land \Ms^2(\Yrf, \mathtt{justifies}) \land{}
    \\
    &\quad\forall y\, \Bigl(
      \bigl(\mathtt{read}(y) \land X(y)\bigr)
      \limp
      \exists x\,\bigl(\mathtt{write}(x) \land X(x) \land \Yrf(x,y)\bigr)
    \Bigr)
  \end{aligned}
  \right.
\end{align}
\end{small}
When $X$~is a potential execution and $\Yco$~is a potential coherence-order relation, the formula $\Mco(X,\Yco)$ requires that the writes in~$X$ for the same location includes some total order.
Because of the later condition that $\Mr(\Yco,\Yrf)$ is acyclic, $\Yco$~is in fact required to be a total order per location.
When $X$~is a potential execution and $\Yrf$~is a potential reads-from relation, the formula $\Mrf(X,\Yrf)$ requires that $\Yrf$~is injective, is a subset of $\mathtt{justifies}$, and relates all the reads in~$X$ to some write in~$X$.

The auxiliary relation $\Mr(\Yco,\Yrf)$ is the union of strict program-order~($<$), reads-from~($\Yrf$), coherence-order~($\Yco$), and the from-reads relation:
\begin{align}
  \Mr(\Yco,\Yrf)(y,z) \;&\defeq\;
    (y < z)
    \lor
    \Yco(y,z)
    \lor
    \Yrf(y,z)
    \lor
    \exists x\, \bigl( \Yco(x,z) \land \Yrf(x,y) \bigr)
\end{align}

\subsection{Release--Acquire}\label{secgc} 

The Release--Acquire memory model is similar to sequential consistency but more relaxed.
The structure it operates has the same vocabulary, and the memory model is captured by the formula~$\Mra$, defined as follows:
\begin{align}
  \Mra \;\defeq\; \exists X\Yco\Yrf\, \left(
    \begin{aligned}
    &\Mf(X)
    \land \Mco(X,\Yco)
    \land \Mrf(X,\Yrf)
    \land \Macyclic(\Yco)
    \\&\quad{}\land
      \exists \Yhb\,
      \left(
      \begin{small}
      \begin{aligned}
      &\Ms^2({<},\Yhb) \land \Ms^2(\Yrf,\Yhb) \land \Mtrans(\Yhb)
      \\&{}\quad
      \land\Mirrefl(\Yhb)
      \land\Mirrefl(\Mseq(\Yco,\Yhb))
      \\&{}\quad
      \land\Mirrefl(\Mseq(\Minv(\Yrf), \Mseq(\Yco, \Yhb)))
      \end{aligned}
      \end{small}
      \right)
    \end{aligned}
  \right)
\end{align}
The existential SO variable $\Yhb$ over-approximates a relation traditionally called happens-before.

\subsection{C++}\label{sec:mm:cpp} 
To capture the C++ model in SO logic, we follow the {\tt .cat} model of Lahav et
al.~\cite{repairingsc}. Their work introduces necessary patches to the model of
the standard~\cite{mathematizing} but also includes fixes and adjustments from
prior work~\cite{overhaulingsc,tamingra}. The model is more nuanced than the SC
and RA models and requires additions to the vocabulary of~$\fA$, but the key
difference is more fundamental. C++ is a \emph{catch-fire} semantics: programs
that exhibit even a single execution with a data race are allowed to do anything
at all, even burst into flames, and this means that they satisfy every expected
outcome. This difference is neatly expressed in SO logic:

\begin{align}
  \Mcpp \;\defeq\; \exists X\Yco\Yrf\, \left(
    \begin{aligned}
    & \Mco(X,\Yco) \land \Mrf(X,\Yrf) \land \Mm(\Yco,\Yrf) \\
    &\quad{}\land (\Mf(X) \lor \Mc(\Yco,\Yrf))
    \end{aligned}
  \right)
\end{align}

The formula reuses $\Mco$, $\Mrf$ and $\Mf(X)$ and includes two new
macros: $\Mm(\Yco,\Yrf)$ and $\Mc(\Yco,\Yrf)$.
$\Mm(\Yco,\Yrf)$ captures the conditions imposed on a valid C++
execution, and is the analogue of the conditions applied in $\Msc$
and $\Mra$.
$\Mc(\Yco,\Yrf)$ holds if there is a race in the execution $X$. Note
that the expected outcome is allowed if $\Mf(X)$ is satisfied or if
there is a race and $\Mc(\Yco,\Yrf)$ is true.

%
%

\subsection{Jeffrey--Riely}\label{sec:mm:jr} 

The J+R memory model is captured by a sentence $\Mjr_n$, parametrised by an integer~$n$.
Unlike the formulas we saw before, $\Mjr_n$ makes use of three levels of quantifiers ($\exists\forall\exists$), putting it on the third level of the polynomial hierarchy.
We begin by lifting%
\footnote{%
  Our definition of $\Mj$ is different from the original one~\cite{Jeffrey:2016}: we require that only new reads are justified, by including the conjunct~$\lnot P(y)$.
  Without this modification, our solver's results disagree with the hand-calculations reported by Jeffrey and Riely; with this modification, the results agree.}
  $\mathtt{justifies}$ from events to sets of events $P$~and~$Q$:
\begin{align}
  \Mj(P,Q) &\defeq
  \forall y\,
    \left(
    \begin{aligned}
      &\bigl( \lnot P(y) \land Q(y) \land \mathtt{read}(y)  \bigr)\\
      &\quad\limp
      \exists x\, \bigl(
        P(x) \land \mathtt{write}(y) \land \mathtt{justifies}(x,y)
      \bigr)
    \end{aligned}
    \right)
\\
  \Maj(P,Q) &\defeq
    \Mj(P,Q) \land \Ms^1(P,Q) \land \Mv(P) \land \Mv(Q)
\end{align}
We read $\Mj$ as `justifies', and $\Maj$ as `always justifies'.
Next, we define what Jeffrey and Riely call `always eventually justify'
\begin{align}
  \Maej_n(P,Q) &\defeq
\left\{
\begin{aligned}
    &\Ms^1(P,Q) \land \Mv(P) \land \Mv(Q) \land {} \\
    &\forall X\,
      \Bigl( \Mtc_n(\Maj)(P,X) \limp
        \exists Y\,\bigl(\Mtc_n(\Maj)(X,Y) \land \Mj(Y,Q)\bigr) \Bigr)
\end{aligned}
\right.
\end{align}
The size of the formula $\Mtc_n(\Maej_m)(P,Q)$ we defined above is $\Theta(mn)$.
In particular, it is bounded.
Finally, we let%
  \footnote{The symbol $\emptyset$ denotes the empty unary relation, as expected.}
\begin{align}
  \Mjr_n \defeq
    \exists X\,
      \bigl( \Mtc_n(\Maej_n)(\emptyset,X)
      \land \Mf(X) \bigr)
\end{align}
and ask solve the model checking problem $\fA \models \Mjr_n$.
Since the formulas above are in MSO, it is sufficient to pick $n \defeq 2^{|A|}$.
Since all bounded transitive closures include the subset relation, they are monotonic, and it suffices, in fact, to pick $n \defeq |A|$.
\RG{TODO: prove.}
For actual solving, we will use this observation.

\section{Encoding in QBF}\label{sec:qbf} 

In the previous section, we saw that deciding whether a given program behaviour
is allowed by a given memory model can often be expressed naturally as a model
checking problem $\fA \models \phi$ in second-order logic.  Now we want to solve
such problems. We do not use existing model finders and
solvers~\cite{blanchette-itp10,brown-ijcar12,ic3+ia}: we find those for
first-order logic are efficient, but not expressive enough; whereas those for
higher-order logic are expressive but not efficient.  As a middle road, we
reduce the model-checking problem in second-order logic to checking the validity
of a QBF\null.  This reduction is simple and natural, and it lets us profit from
the recent improvements in QBF solving.
We first define QBF~(\autoref{sec:qbf:def}) and then present the translation from SO to QBF~(\autoref{sec:qbf:translation}).

\subsection{Quantified Boolean Formulas}\label{sec:qbf:def} 

QBF can be seen as a restriction of second-order logic:
(i)~we banish second-order quantifiers from formulas; and
(ii)~we fix the structure.
The universe contains two elements, $0^\fA$~and~$1^\fA$, denoted by the constant symbols $0$~and~$1$, respectively.
There is a unique relational symbol~$T$ which denotes the relation $\{1^\fA\}$.
We denote this fixed structure by~$\sQbf$.
Instead of writing $T(0)$ and $T(1)$ we abuse notation, as is common, and write $0$~and~$1$.

\subsection{Translation from SO to QBF}\label{sec:qbf:translation} 

Given a structure~$\fA$ and an SO sentence~$\phi$, we will construct a QBF sentence $\eval{\fA\models\phi}$ such that $\fA\models\phi$ holds if and only if $\sQbf\models \eval{\fA\models\phi}$ holds:
\begin{align}
  \eval{\fA \models P(t_1,\ldots,t_k)}_{\gamma,\Gamma} \;&\defeq\;
    \Gamma(P)(\gamma(t_1),\ldots,\gamma(t_k))
  \label{eq:so-to-qbf:atom}
\\
  \eval{\fA \models \phi_1 \lnand \phi_2}_{\gamma,\Gamma} \;&\defeq\;
    \eval{\fA\models\phi_1}_{\gamma,\Gamma}
    \lnand \eval{\fA\models\phi_2}_{\gamma,\Gamma}
\\
  \eval{\fA \models \forall x\, \phi}_{\gamma,\Gamma} \;&\defeq\;
    \bigwedge_{i=1}^n \eval{\fA \models \phi}_{\gamma[x\mapsto a_i^\fA],\Gamma}
  \label{eq:so-to-qbf:all:fo}
\\
  \eval{\fA \models \exists x\, \phi}_{\gamma,\Gamma} \;&\defeq\;
    \bigvee_{i=1}^n \eval{\fA \models \phi}_{\gamma[x\mapsto a_i^\fA],\Gamma}
  \label{eq:so-to-qbf:any:fo}
\\
  \eval{\fA \models \forall X^k\, \phi}_{\gamma,\Gamma} \;&\defeq\;
    \forall\vec{\sf x}\,
      \eval{\fA\models\phi}_{\gamma,\Gamma[X^k\mapsto \vec{\sf x}\,]}
  \label{eq:so-to-qbf:all:so}
\\
  \eval{\fA \models \exists X^k\, \phi}_{\gamma,\Gamma} \;&\defeq\;
    \exists\vec{\sf x}\,
      \eval{\fA\models\phi}_{\gamma,\Gamma[X^k\mapsto \vec{\sf x}\,]}
  \label{eq:so-to-qbf:any:so}
\end{align}
As before, $\gamma$~maps first-order variables to universe elements. 
Unlike before, $\Gamma$~maps SO variables $X^k$ to (total) functions from $A^k$ to QBF~terms.
For example, $\Gamma(X^2)(a_1^\fA,a_2^\fA)$ is a QBF term.
As before, we make the convention that the empty first-order environment maps constants to the elements they denote: $\epsilon(a)\defeq a^\fA$.
For the empty SO environment~$E$, we make the following convention:
\begin{align}
  E(R)(\vec{a}) \defeq
  \begin{cases}
  0 & \text{if $\vec{a}\in R^\fA$} \\
  1 & \text{if $\vec{a}\not\in R^\fA$}
  \end{cases}
\end{align}
Above, $0$~and~$1$ are QBF constants, and $\vec{a}\in A^k$ where $k$~is the arity of~$R$.
The notation $\eval{\fA\models\phi}$ is shorthand for $\eval{\fA\models\phi}_{\epsilon,E}$.

In \eqref{eq:so-to-qbf:all:so}~and~\eqref{eq:so-to-qbf:any:so}, SO quantifiers are handled by introducing $|A|^k$ QBF variables~$\vec{\mathsf{x}}$, where $k$~is the arity.
The SO environment $\Gamma$ is extended with a binding from the SO variable~$X^k$ to a bijective function from~$A^k$ to the fresh variables.
In~\eqref{eq:so-to-qbf:atom}, this function is extracted from the environment and applied.
Intuitively, the QBF variable $\vec{\mathsf{x}}(a_1^\fA,\ldots,a_k^\fA)$ tracks whether $(a_1^\fA,\ldots,a_k^\fA)$ belongs to~$X^k$.

In \eqref{eq:so-to-qbf:all:fo}~and~\eqref{eq:so-to-qbf:any:fo}, first-order quantifiers are handled by simply expanding them into corresponding boolean connectives.
This eager expansion is a potential target for optimisation in the future.


\begin{figure}[t]
\noindent\begin{minipage}{\textwidth}
  \begin{minipage}{0.45\textwidth}
    \includegraphics[width=\textwidth]{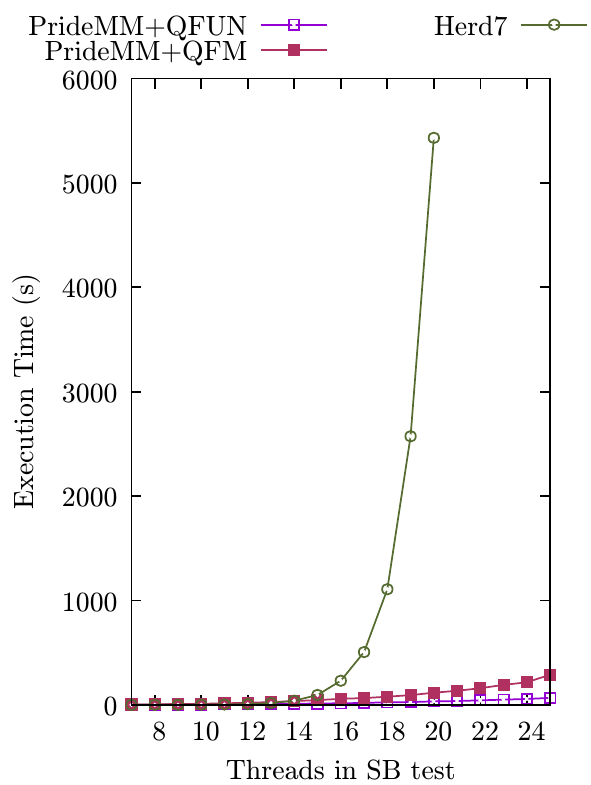}
    \captionof{figure}{\label{fig:comp-herd}Comparison between Herd
      and PrideMM on the store buffer problem.}
  \end{minipage}
  \hfill
  \begin{minipage}{0.52\textwidth}
    \centering
    \begin{small}
    \begin{tabular}{|l|c|c|c|c|}
        \hline
\textbf{Prob.} &   \textbf{SAT} &     \textbf{caqe (s)} &   \textbf{qfun (s)} &   \textbf{qfm (s)} \\\hline
1  &  N  & $\bot$ &  610  &  \textbf{2}  \\
2  &  N  & $\bot$ &  23  &  \textbf{2}  \\
3  &  Y  & $\bot$ & $\bot$ &  \textbf{222}  \\
4  &  Y  & $\bot$ &  \textbf{2}  &  5  \\
5  &  Y  & $\bot$ &  78  &  \textbf{51}  \\
6  &  N  &  5  &  4  &  \textbf{1}  \\
7  &  Y  & $\bot$ &  280  &  \textbf{56}  \\
8  &  N  & $\bot$ &  \textbf{2}  &  \textbf{2}  \\
9  &  N  & $\bot$ &  2  &  \textbf{1}  \\
10  &  Y  & $\bot$ &  36  &  \textbf{10}  \\
11  &  Y  & $\bot$ &  598  &  \textbf{335}  \\
13  &  Y  &  \textbf{1}  &  \textbf{1}  &  \textbf{1}  \\
14  &  Y  & $\bot$ &  \textbf{29}  &  33  \\
15  &  Y  & $\bot$ &  512  &  \textbf{157}  \\
16  &  N  & $\bot$ & $\bot$ &  \textbf{12}  \\
17  &  N  & $\bot$ &  \textbf{39}  &  311  \\
18  &  N  & $\bot$ &  359  &  \textbf{190}  \\
        \hline
\textbf{\#17} &     \textbf{} &     \#2  &  \#15  &  \#17  \\
        \hline
    \end{tabular}
%
    \end{small}
    \captionof{table}{\label{tab:j+r}CPU time for solving the litmus
      tests with J+R model; $\bot$ represents time/mem-out.}
  \end{minipage}

\end{minipage}
\end{figure}
\section{Evaluation} 
\label{sec:evaluation}
The evaluation aims to analyse the performance and correctness of the developed tool.
To this end we included ``tricky'' benchmarks that are studied in the literature and
benchmarks for scaling.  Additionally, various beckends to the presented tool PrideMM are considered.



\paragraph{Solvers.} We evaluate the QBF approach using off-the-shelf
solvers CAQE~\cite{rabe-fmcad15} and QFUN~\cite{janota-aaai18}, the
respective winners of the CNF and non-CNF tracks at 2017's QBFEVAL
competition~\cite{qbfeval17}. Our QBF benchmarks were first produced in
the circuit-like format QCIR~\cite{klieber-bnp16}, natively supported
by QFUN. The inputs to CAQE were produced by converting to CNF through
standard means, followed by a preprocessing step with
bloqqer~\cite{biere-cade11}.

Encouraged by the results of the QBF approach, we have started the
development of a dedicated solver for SO model checking. The solver is
called \emph{QFM} and it accepts as input a structure and an SO formula.
Currently the solver expands all first-order quantifications,
following a similar approach to the translation of
Section~\ref{sec:qbf:translation}.
The QBF problem is then solved using the non-prenex version of the
RAReQS algorithm~\cite{janota-ai16}.
A dedicated SO solver is able to use specialised techniques,
e.g.\ lazily expanding quantifiers. Such techniques present a
particular advantage for universes with large number of elements: the
inherent exponential characteristic of the expansion step will
eventually lead to issues in the translation to QBF.

\paragraph{Instances and memory models.}
In our first set of instances, we simulate a series of n-threaded store-buffering tests
(Figure~\ref{fig:store-buf}) over sequential
consistency~\cite{lamportsc}, and compare the performance of PrideMM
and Herd7~\cite{herd}. The results of this comparison are shown in
Figure~\ref{fig:comp-herd}.
In a second set of instances, we simulate the J+R model on the Java
causality tests~\cite{javamm}.  There are no other tools to benchmark
against; ours is the only simulator for this model. Instead, we
provide a comparative evaluation between our QBF and QFM backends.
In a final set of instances, we simulate a collection of standard
tests taken from the literature on axiomatic memory
models~\cite{DBLP:conf/pldi/SarkarSAMW11}. Each of these completes in
under 6s.

\begin{figure}[t]
  \centering
  \begin{tabular}{c}
    initially ${\tt x_1 = 0, x_2 = 0, \: \hdots \:, x_n = 0}$\\
    \hline
    \begin{tabular}{l @{\;} || @{\;} l || @{\;} l || @{\;} l || l @{\;}}
      ${\tt x_1 = 1}$     & ${\tt x_2 = 1}$   & \hspace{2ex} $\hdots$ \hspace{2ex} & ${\tt x_{n-1} = 1}$   & ${\tt x_n = 1}$ \\
      ${\tt r_1 = x_{n}}$ & ${\tt r_2 = x_1}$ & \hspace{2ex} $\hdots$ \hspace{2ex} & ${\tt r_{n-1} = x_{n-2}}$ & ${\tt r_n = x_{n-1}}$ \\
    \end{tabular} \\
    \hline
    ${\tt r_1 == 0 \wedge r_2 == 0 \wedge \hdots \wedge r_{n-1} == 0 \wedge r_n == 0}$ allowed?
  \end{tabular}
  \caption{\label{fig:store-buf} The store-buffer problem.}
\end{figure}

\paragraph{Discussion of the results.}
Figure~\ref{fig:comp-herd} indicates a stark contrast in the
scalability of the store-buffering problem on PrideMM when compared
with Herd7. PrideMM enables the practical simulation of far larger
tests: 25-thread SB -- with 100 events -- solves in 1 minute.
Axiomatic tests reduce to SAT problems, so one might expect
similar performance from QFUN and QFM, but QFUN has the
more mature implementation.

Table~\ref{tab:j+r} demonstrates the viability of our approach to
simulating the J+R model. QFUN solves all but two instances, whereas
QFM solves all of them, taking no longer than 6~min on any
instance. We found the CNF-based QBF solver CAQE to be inadequate for
these problems. The timeout was set to 30 minutes, and the memory
available was 32GB.
The dedicated SO solver QFM performs better than the off-the-shelf QBF
solver QFUN -- even though they implement the same algorithm. We
attribute this to a more efficient implementation of the expansion of
first-order logic quantifiers (e.g.\ repetition of subformulas is avoided by
hash-consing already during expansion).
Additionally, QFM supports non-prenex input, while QFUN operates on
prenex form.
The satisfiability of each instance matches the expected
results~\cite{Jeffrey:2016}.



\section{Related Work}
Our evaluation was limited to 4 memory models: SC, RA, C++ and
J+R.
Although we have covered a breadth of axiomatic models, there are
several others that fall into the class of the J+R model that we have
not covered, i.e. the promising model of Kang et al.~\cite{promising},
or the model of Pichon--Pharabod and Sewell (P+S)~\cite{bubbly}.
It is clear that Promising and P+S are definable in higher-order logic
and hence in second-order logic, by the standard encoding of
higher-order in second-order (over finite structures).
Moreover, for J+R, we do not show that the model definable directly as
a second-order logic formula~$\phi$, but instead describe it as a
sequence $\{\phi_n\}_{n\ge 0}$ of formulas, one for each universe
size.
Thus, our decision to stay in second-order logic and use parametrised
formulas does not prevent us from representing other models, and
experimental validation indicates that we have found a pragmatic
sweet-spot for simulating this new class of models.

We use Herd as a performance benchmark because it is the predominant
weak-memory modelling tool, but there are
others. CDSChecker~\cite{cdschecker} is a model checker entirely
specialised to the axiomatic model of C++. Memalloy~\cite{memalloy}
uses SAT solvers to model a range of models, but cannot model the J+R
model efficiently.

There are other weak-memory questions that one might seek to answer
automatically beyond simulation: Memalloy~\cite{memalloy} can compare
axiomatic memory models to find programs that act as differentiating
counterexamples, with executions allowed by one and not the
other. Bornholt and Torlak's MemSynth~\cite{memsynth} can synthesise
axiomatic memory models from sets of litmus tests. We choose synthesis
as our task because it is a good starting point with clear utility.


\section{Conclusion} 
This paper presents PrideMM, a tool that vastly exceeds the performance of
Herd, a state-of-the-art simulator for axiomatic concurrency models, and that
simulates one of a new class of models for which previous techniques
do not apply. We argue that for weak-memory model simulation, SO
logic provides a useful balance of expressiveness and performance when
combined with state-of-the-art solvers.

\clearpage
\bibliographystyle{splncs03}
\bibliography{bibliography,solvers}

\begin{thebibliography}{10}
\providecommand{\url}[1]{\texttt{#1}}
\providecommand{\urlprefix}{URL }

\bibitem{gpu}
Alglave, J., Batty, M., Donaldson, A.F., Gopalakrishnan, G., Ketema, J.,
  Poetzl, D., Sorensen, T., Wickerson, J.: {GPU} concurrency: Weak behaviours
  and programming assumptions. In: Proceedings of the Twentieth International
  Conference on Architectural Support for Programming Languages and Operating
  Systems, {ASPLOS} '15, Istanbul, Turkey, March 14-18, 2015. pp. 577--591
  (2015), \url{http://doi.acm.org/10.1145/2694344.2694391}

\bibitem{lisa}
Alglave, J., Cousot, P.: Syntax and analytic semantics of {LISA}.
  \url{https://arxiv.org/abs/1608.06583} (2016)

\bibitem{cat}
Alglave, J., Cousot, P., Maranget, L.: Syntax and analytic semantics of the
  weak consistency model specification language {CAT}.
  \url{https://arxiv.org/abs/1608.07531} (2016)

\bibitem{powerbug}
Alglave, J., Maranget, L., Sarkar, S., Sewell, P.: Fences in weak memory models
  (extended version). Formal Methods in System Design  40(2),  170--205 (2012),
  \url{https://doi.org/10.1007/s10703-011-0135-z}

\bibitem{herd}
Alglave, J., Maranget, L., Tautschnig, M.: Herding cats: Modelling, simulation,
  testing, and data mining for weak memory. {ACM} Trans. Program. Lang. Syst.
  36(2),  7:1--7:74 (2014), \url{http://doi.acm.org/10.1145/2627752}

\bibitem{selman-aaai05}
Ans{\'o}tegui, C., Gomes, C.P., Selman, B.: The {Achilles'} heel of {QBF}. In:
  AAAI. pp. 275--281 (2005)

\bibitem{BalabanovSAT16}
Balabanov, V., Jiang, J.R., Mishchenko, A., Scholl, C.: Clauses versus gates in
  {CEGAR-Based} {2QBF} solving. In: Beyond NP, {AAAI} Workshop (2016)

\bibitem{overhaulingsc}
Batty, M., Donaldson, A.F., Wickerson, J.: Overhauling {SC} atomics in {C11}
  and opencl. In: Proceedings of the 43rd Annual {ACM} {SIGPLAN-SIGACT}
  Symposium on Principles of Programming Languages, {POPL} 2016, St.
  Petersburg, FL, USA, January 20 - 22, 2016. pp. 634--648 (2016),
  \url{http://doi.acm.org/10.1145/2837614.2837637}

\bibitem{thinairproblem}
Batty, M., Memarian, K., Nienhuis, K., Pichon{-}Pharabod, J., Sewell, P.: The
  problem of programming language concurrency semantics. In: Programming
  Languages and Systems - 24th European Symposium on Programming, {ESOP} 2015,
  Held as Part of the European Joint Conferences on Theory and Practice of
  Software, {ETAPS} 2015, London, UK, April 11-18, 2015. Proceedings. pp.
  283--307 (2015), \url{https://doi.org/10.1007/978-3-662-46669-8_12}

\bibitem{mathematizing}
Batty, M., Owens, S., Sarkar, S., Sewell, P., Weber, T.: Mathematizing {C++}
  concurrency. In: Proceedings of the 38th {ACM} {SIGPLAN-SIGACT} Symposium on
  Principles of Programming Languages, {POPL} 2011, Austin, TX, USA, January
  26-28, 2011. pp. 55--66 (2011),
  \url{http://doi.acm.org/10.1145/1926385.1926394}

\bibitem{SatHandbook}
Biere, A., Heule, M., van Maaren, H., Walsh, T. (eds.): Handbook of
  Satisfiability, Frontiers in Artificial Intelligence and Applications, vol.
  185. IOS Press (2009)

\bibitem{biere-cade11}
Biere, A., Lonsing, F., Seidl, M.: Blocked clause elimination for {QBF}. In:
  The 23rd International Conference on Automated Deduction {CADE} (2011)

\bibitem{blanchette-itp10}
Blanchette, J.C., Nipkow, T.: Nitpick: {A} counterexample generator for
  higher-order logic based on a relational model finder. In: Interactive
  Theorem Proving, First International Conference ({ITP}). pp. 131--146 (2010),
  \url{https://doi.org/10.1007/978-3-642-14052-5_11}

\bibitem{memsynth}
Bornholt, J., Torlak, E.: Synthesizing memory models from framework sketches
  and litmus tests. In: Proceedings of the 38th {ACM} {SIGPLAN} Conference on
  Programming Language Design and Implementation, {PLDI} 2017, Barcelona,
  Spain, June 18-23, 2017. pp. 467--481 (2017),
  \url{http://doi.acm.org/10.1145/3062341.3062353}

\bibitem{brown-ijcar12}
Brown, C.E.: Satallax: An automatic higher-order prover. In: Automated
  Reasoning - 6th International Joint Conference ({IJCAR}). pp. 111--117
  (2012), \url{https://doi.org/10.1007/978-3-642-31365-3_11}

\bibitem{ic3+ia}
Cimatti, A., Griggio, A., Mover, S., Tonetta, S.: {IC3} modulo theories via
  implicit predicate abstraction. In: TACAS (2014)

\bibitem{GoultiaevaB10}
Goultiaeva, A., Bacchus, F.: Exploiting {QBF} duality on a circuit
  representation. In: AAAI (2010)

\bibitem{goultiaeva-date13}
Goultiaeva, A., Seidl, M., Biere, A.: Bridging the gap between dual propagation
  and {CNF-based QBF} solving. In: {DATE}. pp. 811--814 (2013)

\bibitem{isamem}
Gray, K.E., Kerneis, G., Mulligan, D.P., Pulte, C., Sarkar, S., Sewell, P.: An
  integrated concurrency and core-isa architectural envelope definition, and
  test oracle, for {IBM} {POWER} multiprocessors. In: Proceedings of the 48th
  International Symposium on Microarchitecture, {MICRO} 2015, Waikiki, HI, USA,
  December 5-9, 2015. pp. 635--646 (2015),
  \url{http://doi.acm.org/10.1145/2830772.2830775}

\bibitem{janota-sat12}
Janota, M., Klieber, W., Marques-Silva, J., Clarke, E.M.: Solving {QBF} with
  counterexample guided refinement. In: SAT. pp. 114--128 (2012)

\bibitem{janota-epia17}
Janota, M., Marques{-}Silva, J.: An {Achilles'} heel of term-resolution. In:
  {EPIA} Conference on Artificial Intelligence. pp. 670--680 (2017)

\bibitem{janota-aaai18}
Janota, M.: Towards generalization in {QBF} solving via machine learning. In:
  {AAAI} Conference on Artificial Intelligence (2018)

\bibitem{acyclicity}
Janota, M., Grigore, R., Manquinho, V.: On the quest for an acyclic graph. In:
  RCRA (2017)

\bibitem{janota-ai16}
Janota, M., Klieber, W., Marques-Silva, J., Clarke, E.: Solving {QBF} with
  counterexample guided refinement. Artificial Intelligence  234,  1--25 (2016)

\bibitem{janota-sat11}
Janota, M., Marques-Silva, J.: Abstraction-based algorithm for {2QBF}. In: SAT.
  pp. 230--244 (2011)

\bibitem{janota-ijcai15}
Janota, M., Marques{-}Silva, J.: Solving {QBF} by clause selection. In:
  International Joint Conference on Artificial Intelligence (IJCAI) (2015)

\bibitem{Jeffrey:2016}
Jeffrey, A., Riely, J.: On thin air reads towards an event structures model of
  relaxed memory. In: Proceedings of the 31st Annual ACM/IEEE Symposium on
  Logic in Computer Science. pp. 759--767. LICS '16, ACM, New York, NY, USA
  (2016), \url{http://doi.acm.org/10.1145/2933575.2934536}

\bibitem{klieber-bnp16}
Jordan, C., Klieber, W., Seidl, M.: {Non-CNF} {QBF} solving with {QCIR}. In:
  AAAI Workshop: Beyond NP. AAAI Workshops, vol. WS-16-05. AAAI Press (2016)

\bibitem{promising}
Kang, J., Hur, C., Lahav, O., Vafeiadis, V., Dreyer, D.: A promising semantics
  for relaxed-memory concurrency. In: Proceedings of the 44th {ACM} {SIGPLAN}
  Symposium on Principles of Programming Languages, {POPL} 2017, Paris, France,
  January 18-20, 2017. pp. 175--189 (2017),
  \url{http://dl.acm.org/citation.cfm?id=3009850}

\bibitem{klieber-sat10}
Klieber, W., Sapra, S., Gao, S., Clarke, E.M.: A non-prenex, non-clausal {QBF}
  solver with game-state learning. In: SAT (2010)

\bibitem{tamingra}
Lahav, O., Giannarakis, N., Vafeiadis, V.: Taming release-acquire consistency.
  In: Proceedings of the 43rd Annual {ACM} {SIGPLAN-SIGACT} Symposium on
  Principles of Programming Languages, {POPL} 2016, St. Petersburg, FL, USA,
  January 20 - 22, 2016. pp. 649--662 (2016),
  \url{http://doi.acm.org/10.1145/2837614.2837643}

\bibitem{repairingsc}
Lahav, O., Vafeiadis, V., Kang, J., Hur, C., Dreyer, D.: Repairing sequential
  consistency in {C/C++11}. In: Proceedings of the 38th {ACM} {SIGPLAN}
  Conference on Programming Language Design and Implementation, {PLDI} 2017,
  Barcelona, Spain, June 18-23, 2017. pp. 618--632 (2017),
  \url{http://doi.acm.org/10.1145/3062341.3062352}

\bibitem{lamportsc}
Lamport, L.: How to make a multiprocessor computer that correctly executes
  multiprocess programs. {IEEE} Trans. Computers  28(9),  690--691 (1979),
  \url{https://doi.org/10.1109/TC.1979.1675439}

\bibitem{Libkin:2004}
Libkin, L.: Elements of Finite Model Theory. Springer (2004)

\bibitem{LonsingSAT16}
Lonsing, F., Egly, U., Seidl, M.: {Q}-resolution with generalized axioms. In:
  Theory and Applications of Satisfiability Testing - {SAT}. pp. 435--452
  (2016)

\bibitem{javamm}
Manson, J., Pugh, W., Adve, S.V.: The java memory model. In: Proceedings of the
  32nd {ACM} {SIGPLAN-SIGACT} Symposium on Principles of Programming Languages,
  {POPL} 2005, Long Beach, California, USA, January 12-14, 2005. pp. 378--391
  (2005), \url{http://doi.acm.org/10.1145/1040305.1040336}

\bibitem{marin-fi16}
Marin, P., Narizzano, M., Pulina, L., Tacchella, A., Giunchiglia, E.: Twelve
  years of {QBF} evaluations: {QSAT} is {PSPACE}-hard and it shows. Fundam.
  Inform.  149(1-2),  133--158 (2016),
  \url{https://doi.org/10.3233/FI-2016-1445}

\bibitem{silva99}
Marques-Silva, J.P., Sakallah, K.A.: {GRASP}: A search algorithm for
  propositional satisfiability. IEEE Transactions on Computers  48(5),
  506--521 (1999)

\bibitem{gccbugs}
Morisset, R., Pawan, P., Nardelli, F.Z.: Compiler testing via a theory of sound
  optimisations in the {C11/C++11} memory model. In: {ACM} {SIGPLAN} Conference
  on Programming Language Design and Implementation, {PLDI} '13, Seattle, WA,
  USA, June 16-19, 2013. pp. 187--196 (2013),
  \url{http://doi.acm.org/10.1145/2491956.2491967}

\bibitem{cdschecker}
Norris, B., Demsky, B.: A practical approach for model checking c/c++11 code.
  ACM Trans. Program. Lang. Syst.  38(3),  10:1--10:51 (May 2016),
  \url{http://doi.acm.org/10.1145/2806886}

\bibitem{peitl-sat17}
Peitl, T., Slivovsky, F., Szeider, S.: Dependency learning for {QBF}. In:
  Theory and Applications of Satisfiability Testing - ({SAT}). pp. 298--313
  (2017), \url{https://doi.org/10.1007/978-3-319-66263-3_19}

\bibitem{bubbly}
Pichon{-}Pharabod, J., Sewell, P.: A concurrency semantics for relaxed atomics
  that permits optimisation and avoids thin-air executions. In: Proceedings of
  the 43rd Annual {ACM} {SIGPLAN-SIGACT} Symposium on Principles of Programming
  Languages, {POPL} 2016, St. Petersburg, FL, USA, January 20 - 22, 2016. pp.
  622--633 (2016), \url{http://doi.acm.org/10.1145/2837614.2837616}

\bibitem{qbfeval}
{QBF Eval}, \url{http://www.qbflib.org/index_eval.php}

\bibitem{qbfeval17}
{QBF Eval 2017}, \url{http://www.qbflib.org/event_page.php?year=2017}

\bibitem{rabe-sat16}
Rabe, M.N., Seshia, S.A.: Incremental determinization. In: Theory and
  Applications of Satisfiability Testing - {SAT}. pp. 375--392 (2016)

\bibitem{rabe-fmcad15}
Rabe, M.N., Tentrup, L.: {CAQE:} {A} certifying {QBF} solver. In: Formal
  Methods in Computer-Aided Design, {FMCAD}. pp. 136--143 (2015)

\bibitem{malik2qbf}
Ranjan, D.P., Tang, D., Malik, S.: A comparative study of {2QBF} algorithms.
  In: SAT. pp. 292--305 (2004)

\bibitem{DBLP:conf/pldi/SarkarSAMW11}
Sarkar, S., Sewell, P., Alglave, J., Maranget, L., Williams, D.: Understanding
  {POWER} multiprocessors. In: Proceedings of the 32nd {ACM} {SIGPLAN}
  Conference on Programming Language Design and Implementation, {PLDI} 2011,
  San Jose, CA, USA, June 4-8, 2011. pp. 175--186 (2011),
  \url{http://doi.acm.org/10.1145/1993498.1993520}

\bibitem{tentrup-sat16}
Tentrup, L.: Non-prenex {QBF} solving using abstraction. In: Theory and
  Applications of Satisfiability Testing (SAT). pp. 393--401 (2016)

\bibitem{Gelder-cp13}
{Van Gelder}, A.: Primal and dual encoding from applications into quantified
  boolean formulas. In: {CP}. pp. 694--707 (2013)

\bibitem{hotspotunsound}
\v{S}ev\v{c}{\'{\i}}k, J., Aspinall, D.: On validity of program transformations
  in the {Java} memory model. In: {ECOOP} 2008 - Object-Oriented Programming,
  22nd European Conference, Paphos, Cyprus, July 7-11, 2008, Proceedings. pp.
  27--51 (2008), \url{https://doi.org/10.1007/978-3-540-70592-5_3}

\bibitem{amdbug}
Wickerson, J., Batty, M., Beckmann, B.M., Donaldson, A.F.: Remote-scope
  promotion: clarified, rectified, and verified. In: Proceedings of the 2015
  {ACM} {SIGPLAN} International Conference on Object-Oriented Programming,
  Systems, Languages, and Applications, {OOPSLA} 2015, part of {SPLASH} 2015,
  Pittsburgh, PA, USA, October 25-30, 2015. pp. 731--747 (2015),
  \url{http://doi.acm.org/10.1145/2814270.2814283}

\bibitem{memalloy}
Wickerson, J., Batty, M., Sorensen, T., Constantinides, G.A.: Automatically
  comparing memory consistency models. In: Proceedings of the 44th {ACM}
  {SIGPLAN} Symposium on Principles of Programming Languages, {POPL} 2017,
  Paris, France, January 18-20, 2017. pp. 190--204 (2017),
  \url{http://dl.acm.org/citation.cfm?id=3009838}

\bibitem{Winskel:1987}
Winskel, G.: Event structures, pp. 325--392. Springer Berlin Heidelberg,
  Berlin, Heidelberg (1987), \url{https://doi.org/10.1007/3-540-17906-2_31}

\bibitem{zhang-aaai06}
Zhang, L.: Solving {QBF} by combining conjunctive and disjunctive normal forms.
  In: AAAI (2006)

\bibitem{zhang-iccad02}
Zhang, L., Malik, S.: Conflict driven learning in a quantified {Boolean}
  satisfiability solver. In: International Conference On Computer Aided Design
  (ICCAD). pp. 442--449 (2002)

\end{thebibliography}
\end{document}